\documentstyle[12pt]{article}
\begin{document}
\centerline{\bf Hong et al. reply to Canul-Chay et al.}
\vskip 1.0cm 
\centerline{ Daniel C. Hong and Paul V. Quinn}
\centerline{ Physics, Lewis Laboratory, 
Lehigh University, Bethlehem, Pennsylvania 18015}
\centerline{Stefan Luding }
\centerline{Particle Technology, Particle Technology, DelftChemTech Julianalaan 136}
\centerline{2628 BL Delft, The Netherlands  }
\vskip 1.0cm
Canul-Chay et al. [1] have conducted the segregation experiment of binary 
granular mixtures in a bed and subjected the bed to vibration.  They report 
that the reverse Brazil-Nut phenomenon (RBNP) was never observed in their 
experiments and thus conclude that such does not exist.  However, we have  
clearly demonstrated in [2] by Molecular Dynamics simulations that the 
reverse Brazil-Nut problem not only exists, but can be determined from the 
solid-liquid  phase boundary by a scaling theory for the crossover from 
RBNP.  Hence,  instead of disputing Canul-Chay et al.'s sound experimental 
results [1], in this  reply we want to draw the reader's attention to the 
experimental set up and the  interpretation of the experimental results.  

First, the system considered in [1] is definitely not the same as ours, because  the vibrating bed exhibits a temperature gradient along the vertical axis.  In our  system, the mixture is in contact with the thermal reservoir at $T$ (this is a  global temperature).  Each particle is kicked by random noise such that the  average kinetic energy $m\langle v^2 \rangle /2=T$.  If the reverse Brazil-Nut  phenomenon cannot be found in such a vibrating system, that might teach us a  lot about the sensitivity of granular matter to details of the boundary conditions  (b.c.) - and, on the other hand, if the RBNP is not found for this b.c. it does not  exclude the possibility that it can be found for others.  

Second, in order to observe the reverse Brazil-Nut problem as obtained in [2], it  might be essential that the bed be subjected to a constant global temperature.  In any case, however, it should be essential to quench the system at a low  enough temperature. There may be a way to define a global temperature for the  vibrating bed with a weak deviation from the constant temperature assumption  [3-7]. We now briefly describe some attempts. The density profile of a weakly  excited mono-disperse granular material [4] can be either fit by a Fermi function  [3] (Fig 1 of Ref. [3]), by the Enskog profile plus a rectangle for high densities  [5], or by the integration of the so-called global equation of state (GEQS) [6].  In  Ref. [5], we have determined the condensation temperature $T_c$ as the point  where the sum rule breaks down in the Enskog theory and then used this  temperature to fit the density profile by the rectangle plus the Enskog profile of  the experimental data of [4], and then extracted the global thermal temperature  of the bed.  In Ref. [6], the global equation of state was obtained from  independent, homogeneous numerical simulations and, after integration,  leads to a perfect agreement with the experimental data of the density profile,  without a fit-parameter, from which the temperature for the solid-liquid  condensation can also be extracted. This procedure can be repeated for other  species, and this way two condensation temperatures, say $T(A)$ and $T!(B)$  for two species A and B, are determined.  

Third, once the condensation temperatures $T(A)$ and $T(B)$ for the two  species A and B and the global thermal temperature of the vibrating bed are  determined, we now quench the system to a temperature $T$ that is between $T(A)$  and $T(B)$.  In this case, the system might exhibit the crossover from the Brazil-Nut to reverse Brazil-Nut problem as predicted in [2].  If one still does not observe the reverse Brazil-Nut problem, then one should not but conclude that the vibrating bed may be different from our Molecular Dynamics set up, where the bed is subjected to a global temperature.  One has to think through how to achieve the reverse Brazil-Nut problem for the vibrating bed or learn from this difference in phenomenology what causes the different behavior. 
\vskip 0.2 true cm
\noindent Daniel C. Hong and Paul V. Quinn
\newline
Department of Physics, Lewis Laboratory
\newline
Lehigh University
\newline
Bethlehem, PA 18015  
\vskip 0.2 true cm
\noindent Stefan Luding 
\newline
Particle Technology 
\newline
DelftChemTech Julianalaan 136
\newline
2628 BL Delft, The Netherlands  
\vskip 0.2 true cm
\noindent References  
\vskip 0.2 true cm
\noindent [1] Canul-Chay et al., preceding comment. 
\newline
[2] D. C. Hong, P. V. Quinn, and S. Luding, Phys Rev. Lett. {\bf 86}, 3423, (2001). 
\newline
[3] H. Hayakawa and D. C. Hong, Phys. Rev. Lett., {\bf 78}, 2764, (1997).  
\newline
[4] E. Clement and J. Rajchenbach, Europhys. Lett. {\bf 16}, 1333, (1991).  
\newline
[5] D.C. Hong, Physica A {\bf 271}, 192 (1999).  
\newline
[6] S. Luding, Phys. Rev. E. {\bf 63}, 042201, 2001.  
\newline
[7] P. V. Quinn and D. C. Hong, Phys. Rev. E. {\bf 62}, 8295 (2000). 
\end{document}